\documentclass[showkeys,aps]{revtex4}
\usepackage{amsmath}

\begin{document}

\title{Two simple families of exact inhomogeneous stiff cosmologies}

\author{Guillermo A. Gonz\'{a}lez\footnote{e-mail: gonzalez@gag-girg-uis.net}}
\author{Fabio D. Lora\footnote{e-mail: fadu31@gmail.com}}
\author{Jenrry A. Jaimes\footnote{e-mail: jenrry@gmail.com}}

\affiliation{Grupo de Investigaci\'on en Relatividad y Gravitaci\'on \\
Escuela de F\'{\i}sica, Universidad Industrial de Santander \\
A.A. 678, Bucaramanga, Colombia}

\begin{abstract}

Two families of exact simple solutions of Einstein field equations for
inhomogeneous stiff cosmologies are presented. The method to obtain the
solutions is based on the introduction of auxiliary functions in order to cast
the Einstein equations in such a way that can be explicitly integrated. Now,
despite of the equations are mathematically equivalent to the equations
obtained when the source of matter is a scalar field, is worth to mention that
the source that we consider is not a scalar field but a perfect fluid with the
stiff equation of state. The obtained solutions are expressed in terms of
simple functions of the used coordinates and two families of particular
solutions are considered. The geometrical and kinematical properties of the
solutions are then analyzed and the parameters are restricted in order to have
a physically acceptable behavior. The two particular solutions are of the
Petrov type I, the first one being regular everywhere whereas the second one
presents a big-bang singularity. Now, for a particular value of one of the
parameters, the second particular solution is a vacuum solution of the Bianchi
I type that reduces to the Kasner solution.

\end{abstract}


\maketitle

\section{Introduction}

The main motivation for the formulation of the standard physical cosmology, in
which the universe is described as isotropic and completely homogeneous in all
its evolution\cite{peeb1,peeb2,weinb,zelnov}, has been the observable fact that
the present universe seems to be isotropic and spatially homogeneous. However,
as is shown by a series of recent observations\cite{smoot,bennet,sperg}, our
universe certainly is neither exactly homogeneous nor isotropic, and also there
is not sufficient reason to believe that the behavior of its expantion was
regular at early times. Thus, in order to understand the evolution and
large-scale structure of the universe, it is necessary to consider a more
general class of cosmologies obtained by removing the requirement of
homogeneity and isotropy. Accordingly with the above considerations, in the
last two decades there has been an increasing interest in the study of
anisotropic and inhomogeneous cosmologies, as can be see, for instance, in the
references \cite{waimar,wain1,wain2,carcha,wain3,ahzflz}. (See also the
reviews\cite{cachma,bogrma,mac3,mac4,mac5,verd1} for a summary of the main
work).

Now, between the several methods used to study the spatially inhomogeneous
cosmologies, the search for exact solutions of Einstein field equations plays a
specially important role. Indeed, due to the highly nonlinear character of the
field  equations, the knowledge of some exact solutions is crucial for the
understanding of specific qualitative aspects that can be used as a guide for
the study of more general models. Now then, the properties of the exact
solutions must be related and must be compared with the results obtained by
other different methods, such as the methods of dynamical
systems\cite{wainell}, the methods of the theory of perturbations\cite{SKMH} or
the study of the structure and formation of singularities\cite{seno1}.

However, due to the mathematical complexity of the generic inhomogeneous
models, the study of exact inhomogeneous cosmologies have been limited mainly
to the case of spacetimes admitting an Abelian two-parameter group of
isometries, $G_2$, and were initiated with the study of Gowdy of spatially
compact models\cite{gowdy1,gowdy2}. Now, if the metric admits an orthogonally
transitive two-parameter group of isometries, then can be written in a blocks
diagonal form\cite{SKMH}. This kind of metrics admits two spacelike commuting
Killing vector fields and, as can be easily verified\cite{verdaguer}, the
corresponding vacuum Einstein equations leads to a complete integrable system
of partial differential equations.

On the other hand, when we consider the non vacuum Einstein equations, the
mathematical complexity makes it very difficult to find solutions with
reasonably realistic fluids. So, in almost all the works on exact inhomogeneous
cosmologies, the solutions are obtained by taking fluids whose equation of
state is of the form $p = \gamma \rho$ and, in particular, with $\gamma = 1$.
This last case, the stiff fluid state equation, leads to a complete integrable
system of partial differential equations and was considered by
Zeldovich\cite{ZEL} as a good candidate in order to describe the  matter
content of the Universe in its earlier stage.

In agreement with the above considerations, in the present paper we present two
families of simple exact inhomogeneous stiff cosmologies. The method to obtain
the solutions is based on the introduction of auxiliary functions in order to
cast the Einstein equations in such a way that can be explicitly integrated
and  the obtained solutions can be expressed in terms of simple functions of
the used coordinates. Now, it is worth to mention that the equations are
written in a way that are mathematically equivalent to the equations obtained
when the source of matter is a scalar field. However, in spite of this
mathematical equivalence, the source that we are considering is a perfect fluid
with the stiff equation of state and not a scalar field.

The paper is organized as follows. First, in Sec. \ref{sec:ecs}, we present the
Einstein and matter evolution equations and the integration procedure to obtain
an explicit general solution. Then, in Sec. \ref{sec:sols}, we consider two
families of simple particular solutions and analyze their geometrical and
kinematical properties. Finally, in Sec. \ref{sec:disc}, we summarize our main
results. 

\section{The Einstein and Matter Evolution Equations}
\label{sec:ecs}

In order to study inhomogeneous stiff cosmologies, we take as the starting
point the metric tensor as given by the line element\cite{SKMH}
\begin{equation}
ds^2 = e^{-2U} [e^{2\gamma}(-dt^{2} + dr^{2}) + W^{2} dx^{2}] + e^{2U} dy^2,
\label{eq:1}
\end{equation}
where $U$, $\gamma$ and $W$ are functions of $r$ and $t$ only. We also consider
as the matter contents a perfect fluid with the stiff  equation of state $p
=\rho$, whose energy-momentum tensor can be written as
\begin{equation}
T_{\alpha\beta} = \rho(2 u_{\alpha} u_{\beta} + g_{\alpha\beta}). \label{eq:4}
\end{equation}
With the above choices, the Einsten equations can be cast as
\begin{equation}
R_{\alpha\beta} = 2 \rho u_{\alpha} u_{\beta},  \label{eq:3a}
\end{equation}
whereas the matter evolution equations can be obatined, from the conservation
law
\begin{equation}
{T^{\alpha\beta}}_{;\beta} = 0,\label{eq:3b}
\end{equation}
by projecting it along the temporal and spatial directions. In order to obtain
the above projections, we contract the equation (\ref{eq:3b}) with the velocity
vector $u^\alpha$ and the ``spatial projection tensor'' $h_{\alpha\beta} =
u_{\alpha} u_{\beta} + g_{\alpha\beta}$, respectively. So we obtain
\begin{eqnarray}
&&\rho_{,\beta} u^{\beta} + 2 \rho {u^{\beta}}_{;\beta} =
0,  \label{eq:5} \\
    \nonumber   \\
&&2 \rho u^{\beta} {u^{\alpha}}_{;\beta} h^{\mu}_{\alpha} +
\rho_{,\beta} g^{\beta\alpha} h^{\mu}_{\alpha} = 0,  \label{eq:6}
\end{eqnarray}
where we use the condition $u_{\alpha} h^{\alpha\beta}=0$.

We now impose the irrotationality condition \cite{taub}
\begin{equation}
u_{\alpha}=\frac{\Phi_{,\alpha}}{(-\Phi_{,\mu}\Phi^{,\mu})^{1/2}},
\label{eq:7}
\end{equation}
so that the equation (\ref{eq:6}) can be cast as
\begin{equation}
\frac{\rho (\Phi_{,\mu} \Phi^{,\mu})_{,\alpha}
h^{\alpha}_{\mu}}{(\Phi_{,\mu} \Phi^{,\mu})} = \rho_{,\alpha}
h^{\alpha}_{\mu}, \label{eq:8}
\end{equation}
which can be identically satisfied if we choose
\cite{Let}
\begin{equation}
\rho = - \frac{F}{2} \Phi_{,\mu}\Phi^{,\mu}, \label{eq:9}
\end{equation}
where $F$ is an arbitrary function of the scalar potential $\Phi$. Now, by
using (\ref{eq:9}), the energy-momentum tensor can be cast as
\begin{equation}
T_{\alpha\beta} = F \left[ \Phi_{,\alpha} \Phi_{,\beta} -
\frac{1}{2} g_{\alpha\beta} \Phi_{,\mu} \Phi^{,\mu}
\right], \label{eq:11}
\end{equation}
in such a way that the Einstein and evolution equations can be written as
\begin{eqnarray}
R_{\alpha\beta} &=& F \Phi_{,\alpha}\Phi_{,\beta},  \label{eq:12} \\
&   &   \nonumber   \\
F {\Phi^{,\alpha}}_{;\alpha} &=& - \frac{F'}{2} \Phi_{,\mu}
\Phi^{,\mu},\label{eq:10}
\end{eqnarray}
where $F' = \frac{\partial F}{\partial \Phi}$.

For the metric (\ref{eq:1}), the equation (\ref{eq:10}) reduces to
\begin{eqnarray}
\Phi_{,rr} - \Phi_{,tt} + W^{-1} (W_{,r}\Phi_{,r}-W_{,t}\Phi_{,t}) +
\frac{F'}{2F} (\Phi^{2}_{,r} - \Phi^{2}_{,t}) = 0 . \label{eq:15}
\end{eqnarray}
In order to solve the above equation, we assume that $\Phi = \Phi(\psi)$, where
$\psi$ is a new scalar potential, in such a way that (\ref{eq:15}) can be
written as
\begin{eqnarray}
\psi_{,rr} - \psi_{,tt} + W^{-1} (W_{,r}\psi_{,r} - W_{,t}\psi_{,t}) +
\left[\frac{\Phi''}{\Phi'} + \frac{F'\Phi'}{2F}\right] (\psi^{2}_{,r} -
\psi^{2}_{,t})=0.\label{eq:16}
\end{eqnarray}
We now choose the functional dependence of $\Phi (\psi)$ in such a way that the
above equation can be linearized. In order to do this, we take the expression in
the square brackets as equal to zero,
\begin{equation}
\frac{\Phi''}{\Phi'} + \frac{F'\Phi'}{2F} = 0, \label{eq:18}
\end{equation}
in such a way that
\begin{equation}
k \psi = \int \sqrt{F} d\Phi, \label{eq:19}
\end{equation}
where $k$ is an arbitrary positive constant.

Now, by using (\ref{eq:19}), is easy to see that, for any arbitrary function $F
( \Phi )$, the energy density $\rho$, the velocity vector $u_\alpha$, the
energy-momentum $T_{\alpha \beta}$ end the Einstein system of equations can be
cast as
\begin{eqnarray}
&&\rho = -  \frac{k^2}{2} \psi_{,\mu} \psi^{,\mu} , \label{eq:21}
\\
&&u_{\alpha} = \frac{\psi_{,\alpha}} {\sqrt{- \psi_{,\mu}
\psi^{,\mu}}},\label{eq:22}
\\
&&T_{\alpha\beta} = k^2  \left\{ \psi_{,\alpha} \psi_{,\beta} -
\frac{1}{2} g_{\alpha\beta} \psi_{,\mu} \psi^{,\mu}
\right\},\label{eq:23}
\\
&&R_{\alpha\beta} = k^2 \psi_{,\alpha} \psi_{,\beta}.\label{eq:24}
\end{eqnarray}
As we can see, if we take $k = 0$, the above system of equations reduce to the
Einstein vacuum equations. Also, it is worth to mention that the equations are
written in a way that are mathematically equivalent to the equations obtained
when the source of matter is a scalar field. However, in spite of this
mathematical equivalence, the source that we are considering is a perfect fluid
with the stiff equation of state and not a scalar field. 

By using the non vanishing  components of the Ricci tensor for the metric
(\ref{eq:1}), the Einstein and evolution equations can be written as the
following system of partial differential equations
\begin{eqnarray}
&&W_{,rr} - W_{,tt} = 0,\label{eq:29}\\
&&\nonumber \\
&&(W\psi_{,r})_{,r} - (W\psi_{,t})_{,t} = 0,\label{eq:28} \\
&&\nonumber\\
&&(WU_{,r})_{,r} - (WU_{,t})_{,t} = 0,\label{eq:27}\\
&&\nonumber\\
&&\gamma_{,t}W_{,r} + \gamma_{,r}W_{,t} =
2WU_{,t}U_{,r} + k^2 W \psi_{,t}\psi_{,r}+w_{,tr}, \label{eq:26} \\
&& \nonumber\\
&&\gamma_{,t}W_{,t} + \gamma_{,r}W_{,r} = W(U_{,t}^{2}+U_{,r}^{2})  +
\frac{1}{2} \left[ k^2 W (\psi^{2}_{,t} + \psi^{2}_{,r}) +
(W_{,tt}+W_{,rr}) \right]. \label{eq:25}
\end{eqnarray}
As we can see, equation (\ref{eq:29}) is the classical one-dimensional wave
equation, whose solutions are well known. On the other hand, from equations 
(\ref{eq:28}) and (\ref{eq:27}) we can see that $U (t,r)$ and $\psi (t,r)$ both
are solutions of the same partial differential equation. So, in order to
simplify the above system of equations, we can take $U (t,r) = \psi (t,r)$.
Finally, the integrability conditions of the overdetermined system
(\ref{eq:26}) - (\ref{eq:25}) are equivalent to the equations (\ref{eq:29}) -
(\ref{eq:27}), guaranting so the existence of solutions. Now, as can be see
from the above system of equations, the stiff fluids equations are easy to
integrate due to the fact that, as a consequence of the stiff equation of state
$p=\rho$, the equations for the metric functions $U$ and $W$ decouple from the
pressure \cite{fernandez}.

In order to solve the system (\ref{eq:29}) -- (\ref{eq:25}), we first consider
solutions of the equation (\ref{eq:29}) of the general form
\begin{equation}
W(r,t) = \Psi(r + t)  +  \Omega(r - t), \label{eq:30}
\end{equation}
where $\Psi$ and $\Omega$ are arbitrary functions. We also consider another,
linearly independent, solution of (\ref{eq:29}) written as
\begin{equation}
V(r,t) = \Psi(r + t)  -  \Omega(r - t). \label{eq:31}
\end{equation}
Now, by using $V(r,t)$ and $W(r,t)$, we define a coordinate transformation
\begin{equation}
(t,r, x,y) \leftrightarrow (V,W,x,y), \label{eq:32}
\end{equation}
in such a way that the line element takes the form
\begin{equation}
ds^2 = e^{-2U}[e^{2\Lambda}(-dV^{2} + dW^{2}) + W^{2}dx^{2}]+
e^{2U}dy^2,\label{eq:33}
\end{equation}
where $\Lambda$ is given by
\begin{equation}
\Lambda = \gamma - \frac{1}{2} \ln (W^{2}_{,r} - W^{2}_{,t}).
\label{eq:34}
\end{equation}
The above coordinate transformation leads the Einstein and evolution equations
to the form
\begin{eqnarray}
 &&  \Lambda,_{W} = q W(\psi_{,V}^{2}  +  \psi_{,W}^{2}),\label{eq:35}\\
&&\nonumber\\ && \Lambda,_{V} = 2q W \psi_{,V} \psi_{,W},\label{eq:36} \\
&&\nonumber\\
 && (W\psi_{,V})_{,V} - (W\psi_{,W})_{,W}  =  0, \label{eq:37}
\end{eqnarray}
where $q = 1 + k^2/2$. As we can see, equation (\ref{eq:37}) is equivalent to
the integrability condition of the overdetermined system (\ref{eq:35}) -
(\ref{eq:36}).

We will now consider some simple solutions of the above system. In order to
do this, we seek for solutions of (\ref{eq:37}) of the form $\psi(W,V)=
A(W)+B(V)$ and obtain
\begin{equation}
\psi = \frac{a_1}{4} \left( W^2 + 2 V^2 \right) + a_2
\ln W + a_3 V , \label{eq:38}
\end{equation}
where $a_1$, $a_2$ and $a_3$ are real arbitrary constants. According with this,
the solution of (\ref{eq:35}) - (\ref{eq:36}) is given by
\begin{eqnarray}
\Lambda &=& \frac{q}{2}\left(a_1^{2}V^2+a_3^2+2a_1 a_3 V + a_1 a_2 \right) W^2
\nonumber \\
&& + \frac{q a_1^{2}}{16}W^4  + qa_2^2 \ln W +qa_1 a_2 V^2 + 2qa_2 a_3 V.
\label{eq:39}
\end{eqnarray}
Now, by choosing some particular solutions of (\ref{eq:30}) - (\ref{eq:31}), we
can obtain many different families of solutions for the full system
(\ref{eq:29}) - (\ref{eq:25}). So, in the next section, we will present to
simple families of solutions obtained by means of a particularly simple form of
the functions $\Psi$ and $\Omega$.

\section{Two simple families of particular solutions}
\label{sec:sols}

In order to obtain simple particular solutions, we will consider the two simple
functions
\begin{eqnarray}
f_1 (r,t)  &=& \frac{r + t}{2}, \label{eq:f1} \\
&&	\nonumber	\\
f_2(r,t) &=& \frac{r - t}{2} , \label{eq:f2}
\end{eqnarray}
and will take $\Psi$ and $\Omega$ as defined by
\begin{eqnarray}
\Psi (r,t) &=& f_1 (r,t), \label{eq:psi1} \\
&&	\nonumber	\\
\Omega (r,t) &=& f_2(r,t), \label{eq:ome1}
\end{eqnarray}
for the first family of solutions, and
\begin{eqnarray}
\Psi (r,t) &=& f_1 (r,t), \label{eq:psi2} \\
&&	\nonumber	\\
\Omega (r,t) &=& - f_2(r,t), \label{eq:ome2}
\end{eqnarray}
for the second family of solutions.

\subsection{The first family of solutions}

By taking the first family of solutions, we obtain  for the metric functions
the expressions
\begin{eqnarray}
W(r,t) &=& r, \label{eq.w1} \\
\gamma(r,t) &=& \frac{q}{2} \left(a_1^{2}t^2+a_3^2+2a_1 a_3 t + a_1 a_2 \right)
r^2 \nonumber \\
&& + q \frac{a_1^{2}}{16}r^4  + qa_2^2 \ln r  +qa_1 a_2 t^2 + 2qa_2 a_3 t,
\label{eq:41} \\
U(r,t) &=& \frac{a_1}{4} \left( r^2 + 2 t^2 \right) + a_2 \ln r + a_3 t ,
\label{eq:42}
\end{eqnarray}
whereas for the fluid density we obtain the expression
\begin{equation}
\rho = \frac{k^2}{2} \left[ a_1 \left( a_1 t^2 + 2 a_3 t \right) + a_3^2 -
\frac{a_1^2 r^2}{4} - \frac{a_2^2}{r^2} - a_1 a_2 \right] e^{2 (U - \gamma)},
\label{eq:43}
\end{equation}
and for the velocity components the expressions
\begin{eqnarray}
&&u_t = \frac{k}{\sqrt{2 \rho}} \left( a_1 t + a_3 \right), \label{eq:44} \\
&&  \nonumber   \\
&& u_r = \frac{k}{\sqrt{2 \rho}} \left( \frac{a_1 r}{2} + \frac{a_2}{r}
\right). \label{eq:45}
\end{eqnarray}
Now, in order to obtain a physically acceptable distribution of matter, we
require that $\rho \geq 0$ for any value of $r$ and $t$. So, from the
expression (\ref{eq:43}), is easy to see that $\rho$ will be no negative
everywhere only if we take $a_1 = a_2 = 0$. On the other hand, the requirement
that the velocity vector be future oriented for any value of $r$ and $t$ imply
that we must take $a_3 < 0$.

By imposing the above conditions, the line element (\ref{eq:1}) take the
following form
\begin{equation}
ds^2 = e^{-2a_3t} [e^{qa_3^2r^2}(-dt^{2} + dr^{2}) + r^{2} dx^{2}] +
e^{2a_3t} dy^2, \label{eq:line1}
\end{equation}
so that the density is given by the expression
\begin{equation}
\rho = \frac{k^2 a_3^2}{2} e^{ a_3 (2 t - a_3qr^2)} , \label{eq:47}
\end{equation}
whereas the fluid velocity takes the form
\begin{equation}
u^{\alpha} = e^{ a_3 (2 t - a_3qr^2) / 2} ( 1, 0, 0, 0 ), \label{eq:48}
\end{equation}
guaranteing so the comoving nature of the reference frame.

Now, in order to see if the solution has any curvature singularity, we compute
the components of the Weyl tensor in the natural null tetrad of the
metric\cite{Senovilla,verdaguer}, which are given by
\begin{eqnarray}
  \Psi_{0}(t,r) &=& \frac{1}{2} a_3^2 (2 a_3q r-q-2) e^{a_3 (2 t-a_3 q r^2)},  \\
  \Psi_{2}(t,r) &=& -\frac{1}{2} a_3^2 e^{a_3 (2 t-a_3 q r^2)}, \\
  \Psi_{4}(t,r) &=& -\frac{1}{2} a_3^2 (2 a_3q r +q+2) e^{a_3 (2 t-a_3 q
  r^2)},
\end{eqnarray}
so that all of them are regular everywhere. Furthermore, is easy to see that
the Weyl tensor is of Petrov type I.

The kinematical quantities of the metric can be also easily computed. Thus we
obtain the following expressions for the acceleration, the expansion and the
shear of the fluid
\begin{eqnarray}
a_{\alpha} &=& a_{3}^2 q r ( 0, 1, 0,  0 ),\\
\nonumber \\
\theta &=& \vert a_3 \vert e^{-(\vert a_3 \vert t+a_{3}^2 q
r/2)},\label{eq:ex} \\
\nonumber \\
\sigma_{11} &=& \frac{2\vert a_3 \vert}{3}
e^{\frac{a_3}{2}(a_3 q
r^2-2t)},\\
\nonumber \\
\sigma_{22} &=& \frac{2\vert a_3 \vert}{3} r^2
e^{-\frac{a_3}{2}(a_3 q r^2+2t)},\\
\nonumber \\
\sigma_{33} &=& -\frac{4\vert a_3 \vert}{3} r^2
e^{-\frac{a_3}{2}(a_3 q r^2-6t)},\label{eq:shear}
\end{eqnarray}
where all the components have been computed in the natural orthonormal
tetrad of the metric. As we can see the kinematical quantities are
regular everywhere. Also is easy to see that, as $a_3<0$, the components
$\sigma_{11}$ and $\sigma_{22}$ are positive, while
$\sigma_{33}$ is negative.

\subsection{The second family of solutions}

Now, by taking the second family of solutions, we obtain  for the metric
functions the expressions
\begin{eqnarray}
W(r,t) &=& t, \label{eq:w2} \\
\gamma(r,t) &=&  q \left(a_1^{2}r^2+a_3^2+2a_1 a_3 r + a_1 a_2
\right)\frac{t^2}{2} \nonumber \\
&& + q \frac{a_1^{2}}{16}t^4  + qa_2^2 \ln t  +qa_1 a_2 r^2 + 2qa_2 a_3
r,\label{eq:50} \\
U(r,t) &=& \frac{a_1}{4} \left( t^2 + 2 r^2 \right) + a_2
\ln t + a_3 r , \label{eq:51}
\end{eqnarray}
in such a way that the fluid density is given by
\begin{equation}
\rho = \frac{k^2}{2} \left[ \frac{a_1^2 t^2}{4} + \frac{a_2^2}{t^2} + a_1 a_2 -
a_1 \left( a_1 r^2 + 2 a_3 r \right) - a_3^2 \right] e^{2 (U - \gamma)},
\label{eq:52}
\end{equation}
whereas the velocity components are given by
\begin{eqnarray}
&&u_r = \frac{k}{\sqrt{2 \rho}}  \left( a_1 r + a_3 \right), \label{eq:53} \\
&&  \nonumber   \\
&& u_t = \frac{k}{\sqrt{2 \rho}}  \left( \frac{a_1 t}{2} + \frac{a_2}{t}
\right) . \label{eq:54}
\end{eqnarray}

Now, as in the first family of solutions, we require that $\rho \geq 0$ for
any value of $r$ and $t$ in order to obtain a physically acceptable
distribution of matter. From expression (\ref{eq:52}) is easy to see that
$\rho$ will be no negative everywhere only if we take $a_1 = a_3 = 0$, so that
the expression for the density reduces to
\begin{equation}
\rho = \frac{k^2 a_2^2}{2}  t^{ -2 (q a_2^2 - a_2+1)} . \label{eq:55}
\end{equation}
Now, as $(a_2 -
1)/(a_2^2) < 1 < q$, we have an initial singularity and then the density
decreases to zero as $t \rightarrow \infty$. On the other hand, the velocity
vector is given by
\begin{equation}
u^\alpha = t^{a_2 ( 1 - q a_2)} ( 1, 0, 0, 0 ),
\end{equation}
where, in order to have a future oriented timelike vector, we have taken $a_2
< 0$. Also, we can see that the spatial velocity is zero and thus we again
have a comoving reference frame.

The line element can be written as follows
\begin{equation}
ds^2 = t^{-2a_2} [t^{2qa_2^2}(-dt^{2} + dr^{2}) + t^{2} dx^{2}] +
t^{2a_2} dy^2, \label{eq:line2}
\end{equation}
so that, when $q = 1$ (or $k = 0$) we have a vacuum solution of the Bianchi I
type that reduces to the Kasner solution \cite{KAS,LAN}, which can
be written as  \cite{BKL,SKMH}
\begin{equation}
ds^2 = t^{(d^2 - 1)/2} (-dt^{2} + dr^{2}) + t^{1 + d} dx^{2} +
t^{1 - d} dy^2, \label{eq:kas1}
\end{equation}
with the Kasner parameter given by $d = 1 - 2 a_2$. Now, is worth to mention
that another kind o inhomogeneous stiff cosmologies were obtained by Patel and
Dadich \cite{PD} wich also reduce to the Kasner solution. However, in contrast
with the solution here presented, the solution of Patel and Dadich is
singularity free.

Now, in order
to see if the solution has a real initial singularity, we computed the Weyl
tensor in the natural null tetrad of the metric\cite{Senovilla,verdaguer}
and obtain
\begin{eqnarray}
  \Psi_{0}(t,r) &=& \frac{1}{2} a_2 (2 a_2-1) (a_2 q-1) t^{-2 q a_2^2+2 a_2-2},  \\
  \Psi_{2}(t,r) &=& -\frac{1}{2} (a_2-1) a_2 t^{-2 q a_2^2+2 a_2-2}, \\
  \Psi_{4}(t,r) &=& \frac{1}{2} a_2 (2 a_2-1) (a_2 q-1) t^{-2 q a_2^2+2 a_2-2}.
\end{eqnarray}
The scalars constructed from the Ricci and Weyl tensors diverge as
$t\longrightarrow 0$, which corresponds to a big-bang singularity. Also, is
easy to see that in the algebraic classification of the Reimann tensor, the
metric is of Petrov type I.

On the other hand, the kinematical quantities for this model can also be
easily computed and so, by taking  $a_2<0$, we obtain for the non-vanishing
components of the acceleration, expansion and shear of the fluid the
following expressions
\begin{eqnarray}
a_{\alpha} &=& ( 0, 0 , 0,  0 ),\\ \nonumber \\
\theta &=& \frac{(a_{2}^{2} q -a_2+1)}{t^{(a_{2}^{2} q
-a_2+1)}},\label{eq:expantion1} \\
\nonumber \\
\sigma_{11} &=& \frac{(2a_{2}^{2}q-2a_2 -1)t^{(a_{2}^{2}
q-a_2 -1)}}{3} ,\\
\nonumber \\
\sigma_{22} &=& -\frac{(a_{2}^{2}q+2a_2 -2)t^{-(a_{2}^{2} q
+a_2 -1)}}{3},\\
\nonumber \\
\sigma_{33} &=& -\frac{(a_{2}^{2}q-4a_2 +1)t^{-(a_{2}^{2}
q -3a_2 +1)}}{3},\label{eq:shear2}
\end{eqnarray}
where all the components have been computed in the natural orthonormal tetrad
of the metric. It is interesting to see that, as the pressure gradient is zero,
the acceleration is equal to zero and thus the fluid is geodesic. On the other
hand, as $a_2 <0$, all the non-vanishing components of the shear are negative.

\section{Discussion}
\label{sec:disc}

We presented two simple families of exact inhomogeneous stiff cosmologies. The
solutions were obtained by explicitly integrating the Einstein and matter
evolution equations by means of the introduction of an auxiliary function that
leads to a complete integrable system of partial differential equations. Now,
it is worth to mention that the equations were written in a way that are
mathematically equivalent to the equations obtained when the source of matter
is a scalar field. However, despite of this mathematical equivalence, the
source that we consider was not a scalar field but a perfect fluid with the
stiff equation of state.

A general solution was obtained that can be expressed in terms of simple
functions of the used coordinates. Then, two families of particular solutions
were considered, and their geometrical and kinematical properties were analyzed
and the values of the parameters were restricted in order to have a physically
acceptable behavior. The two particular solutions are of the Petrov type I, the
first one being regular everywhere whereas the second one presents a big-bang
singularity.

Now, for a particular value of one of the parameters, the second particular
solution is a vacuum solution of the Bianchi I type that reduces to the Kasner
solution. However, although there are some other exact inhomogeneous stiff
cosmologies that in the vacuum reduce to the Kasner solution, as is the case
with the solutions obtained by Patel and Dadich \cite{PD}, is worth to mention
that this is not a characteristic behavior of this kind of cosmologies. Indeed,
an example of a different behavior is shown by the first particular solution
considered in this paper, which in the vacuum is not of the Bianchi I type and
do not reduces to the kasner solution.

\begin{acknowledgments}

F. D. Lora wants to thank the finantial support from {\it Vicerrector\'ia
Acad\'emica}, Universidad Industrial de Santander.

\end{acknowledgments}

\end{document}